# Antibiotic resistance: Insights from evolution experiments and mathematical modeling

Gabriela Petrungaro[1,*] Yuval Mulla[1,*], and Tobias Bollenbach[1,2,#]

1. Institute for Biological Physics, University of Cologne, 50931 Cologne, Germany
2. Center for Data and Simulation Science, University of Cologne, 50931 Cologne, Germany

* equal contribution

# Corresponding author: t .bollenbach@uni-koeln.de

**Abstract**
Antibiotic resistance is a growing public health problem. To gain a fundamental understanding of resistance evolution, a combination of systematic experimental and theoretical approaches is required. Evolution experiments combined with next-generation sequencing techniques, laboratory automation, and mathematical modeling are enabling the investigation of resistance development at an unprecedented level of detail. Recent work has directly tracked the intricate stochastic dynamics of bacterial populations in which resistant mutants emerge and compete. In addition, new approaches have enabled measuring how prone a large number of genetically perturbed strains are to evolve resistance. Based on advances in quantitative cell physiology, predictive theoretical models of resistance are increasingly being developed. Taken together, a new strategy for observing, predicting, and ultimately controlling resistance evolution is emerging.

**Introduction**
The looming antibiotic resistance crisis has sparked a flurry of research. Approaches from systems and computational biology are increasingly being used to tackle different aspects of this problem. These include big-data approaches and machine learning on clinical data. In particular, a recent study used phenotyping of clinical isolates and patient records to select the best antibiotic treatment [1]. Beyond clinical applications, laboratory experiments in which antibiotic resistance evolves under controlled conditions are crucial for developing mechanistic models and assessing new types of antibiotic treatment protocols, including novel compounds and drug combinations [2–4]. New techniques based on next-generation sequencing are being combined with mathematical models and systematic efforts exploring the effects of genome-wide perturbations on resistance. These efforts have elucidated new ways to predict resistance evolution and even enabled slowing it down.

The experimental study of resistance evolution poses some unique challenges that go beyond those encountered in experiments with clonal organisms. Perhaps most notable is the intrinsic stochasticity of evolution: Each repetition of an evolution experiment can produce different outcomes as the mutations that drive the dynamics occur randomly. While experiments selecting for antibiotic resistance are often remarkably repeatable at the phenotypic and even genotypic levels [5], many replicates of these experiments are required to estimate the probabilities of the various outcomes. Another challenge is that relevant bacterial populations are so large that, even within one, a large number of competing lineages carrying different spontaneous mutations quickly emerge (Fig. 1a) [6,7]. This situation leads to complex population dynamics driven by a combination of stochastic effects and selection. Moreover, the effects of mutations can vary widely between different environments (Fig. 1b). Depending on the system, horizontal gene transfer can further complicate the dynamics [8]. The field is still far from understanding these phenomena in the real world. To make progress toward predictive mathematical models of resistance evolution, new approaches using evolution experiments with reduced complexity have been developed, in particular by tightly controlling population size, selection pressure, and the environment (Fig. 1c).

Here we focus on recent advances in elucidating the dynamics of resistance evolution, made possible by tracking numerous distinct mutant lineages in evolving populations using next-generation sequencing, new high-throughput techniques for massively parallel evolution experiments, and mathematical modeling. Recent reviews already cover other important aspects of this topic, such as the challenge of predicting resistance [9], strategies to counter it using drug combinations [2,3], the role of horizontal gene transfer [8], the resistome [10], the use of WGS in understanding clinical resistance [11], heteroresistance [12], and the challenges of translating basic research on resistance into clinical applications [13].

**From mutation to fixation**

As soon as antibiotics are applied to a large bacterial population, antibiotic-resistant mutants begin to displace the wild type. In bacterial populations larger than $10^6$ cells, such mutants are typically present before any selection is applied or they appear within hours. The population dynamics becomes increasingly complex as more and more resistance mutations appear, resulting in numerous competing lineages (Fig. 1a). Time-resolved whole-genome metagenomic sequencing can be used to study this process [14]: In this approach, the entire DNA of a population is sequenced, so that mutations in different sub-populations can be identified; however, linkage of multiple mutations in the same lineage can only be inferred indirectly. In practice, sequencing depth limits detection to the most prevalent mutations, present at more than 1% in the population; only a small fraction of beneficial mutations reach this limit.

To circumvent this problem, DNA barcodes – short random DNA sequences – were integrated into the chromosomes of a clonal population. By deep sequencing only these barcodes over the course of an evolution experiment, the population size of hundreds of thousands of different lineages can be tracked. As lineages harboring a beneficial mutation increase in frequency, it is possible to infer the distribution of fitness effects [7] despite competition between many lineages carrying different beneficial mutations. Originally developed in yeast [7], this approach was recently adapted to *E. coli* [6,15] and applied to study the dynamics of lineages under weak selection for antibiotic resistance [6]. It was found that a handful of lineages rapidly dominate the population even at antibiotic concentrations far below the minimal inhibitory concentration [6]. This result corroborates that early adaptation to antibiotics is typically dominated by a few mutations with large effect [16]. To track longer-term adaptation to increasing antibiotic concentrations, re-barcoding approaches as recently developed for yeast [17] will be required.

While lineage tracking is a powerful tool to follow the dynamics of lineages consisting of at least a few hundred cells, other techniques such as time-lapse microscopy are needed to follow smaller lineages. Recent experiments combined with stochastic models have directly shown that their population dynamics differ from those of larger lineages in that they are dominated by stochasticity [18,19]. Even highly beneficial mutations are far from certain to establish and take over the population due to effects such as random cell death. A single bacterium with increased resistance may die before it has a chance to proliferate since some antibiotics start killing bacteria well below their minimum inhibitory concentration. As a result, the initial dynamics of mutant establishment are far more random than the process of mutant fixation at higher cell numbers [18,19]. This effect can be amplified by phenotypic delay, in which the protective effect of a resistance mutation takes several generations to manifest, e.g., because the new mutant variant of a protein must first be produced and replace the antibiotic-sensitive ancestral variant [20,21]. The effects of phenotypic delay on mutation rate estimates via Luria-Delbrück fluctuation tests were recently shown using a combination of numerical simulations and experiments [20,21]. Furthermore, as resistance often depends on the breakdown of extracellular antibiotics, some resistance mutations only become effective at higher cell numbers [22]. Therefore, adaptation to antibiotics is governed not only by the distribution of fitness effects of beneficial mutations, but also by the chance of initial survival of the mutants.

**Role of genetic background in resistance evolution**
Beyond these complex dynamics, the course of evolution is determined by the potential of the initial population to increase its resistance. This *resistance evolvability* can vary dramatically among genetically different bacteria [23–27]. A general way bacteria may use to accelerate evolution is to increase the mutation rate in response to stress [28,29]. But the mutation rate is not necessarily the limiting factor for evolving resistance at large population sizes. So how does the genotype of an initially clonal population affect its resistance evolvability?

One promising strategy to address this question is to expose genetically diverse populations to similar antibiotic selection pressures. Recently, different strains from the *E. coli* long-term evolution experiment [30], which had evolved under constant conditions in the absence of antibiotics for 50,000 generations, were used as starting points for selecting spontaneous mutants with increased antibiotic resistance [25]. The acquired resistance levels of the resulting mutants and the selected mutations were dependent on the initial genotype in an idiosyncratic manner [25,31]. These findings illustrate that even relatively modest genetic differences can affect resistance evolvability. A study comparing eight different strains of the bacterial genus *Pseudomonas* found that the strains acquire different levels of resistance to the antibiotic ceftazidime [23]. These differences in resistance evolvability are due to the presence or absence of the transcription factor *ampR* in the respective genomes, which is needed to reach high expression levels of the resistance enzyme beta-lactamase.

To more systematically identify genes that affect resistance evolvability, high-throughput evolution experiments have been developed to screen gene-deletion libraries. Analogous to reverse genetics, which links phenotypes to specific genes, evolving different gene-deletion strains allows systematic identification of how each gene affects resistance evolvability. This strategy is technically challenging as it requires a large number of evolution experiments, which are much longer, require higher replication, and are more difficult to control than basic phenotyping experiments (Box 1). A first high-throughput screen of a gene-deletion library at high antibiotic concentrations identified several gene deletions that increase evolvability, mostly via an increased mutation rate (Box 1a) [27]. Since a high antibiotic concentration was used that rapidly kills susceptible bacteria, this approach was only able to identify gene deletions that rapidly lead to high resistance gains. Deletion strains that have increased resistance a priori have a considerable advantage in such assays, even if their potential to evolve resistance is no higher than for other strains. Therefore, recent studies have used lab automation to evolve hundreds of gene deletion strains with adjustments of the selection pressure (Box 1b,c) [24,26,32].

One study focused on 173 deletions of transcription factor genes and identified specific deletions (including *arcA* and *gutM*) that impede resistance evolution [26]. These effects were found not to be explained by epistatic interactions with resistance mutations and may be partially caused by fitness effects of these gene deletions. Another recent study [24] considered a different set of 98 gene deletions representing diverse other cellular functions in addition to 18 transcription factor genes. Importantly, only deletions causing minor growth defects in the absence of drug were considered and selection pressure and population size were tightly controlled in the experiments (Box 1c) to quantify evolvability despite differences in initial drug sensitivity. Few of the gene deletions were found to accelerate resistance evolution to the antibiotics tetracycline and chloramphenicol, mostly by increasing the mutation rate. In contrast, multiple gene deletions reduced resistance evolvability; these include genes coding for chaperones and lipopolysaccharide biosynthesis enzymes. Several gene deletions related to multidrug-efflux pumps and their regulation drastically lowered evolvability. Deletion of the outer membrane channel *tolC*, a component of several different efflux pumps, even blocked resistance evolution entirely [24].

These effects of gene deletions on resistance evolution are largely due to epistatic interactions: The same spontaneous resistance mutations tend to occur in the relevant deletion strains but with weaker effects [24]. In particular, the *tolC* deletion essentially eliminates the effects of the usual resistance mutations, probably because it massively disrupts the efflux pumps affected by these mutations. Thus, the most common mutational paths to resistance are rendered inaccessible in the absence of *tolC*, which implies the need to explore rare mutations that are less likely to occur or less beneficial. Beyond these effects of specific gene deletions, a global trend has been observed by exploiting the fact that gene deletions generally lead to different initial drug sensitivities [33,34]. Deletion strains that were more sensitive at the beginning of the experiment exhibit greater resistance gains and are able to "catch up" over the course of the evolution experiment [24] – a hallmark of diminishing-returns epistasis, often observed in other evolution experiments [35]. The extent to which these results apply more generally to antibiotics with different modes of action is an interesting question for future research.

The identification of genes that affect resistance evolvability not only provides us with insights to better predict evolution, but also offers a potential way to minimize the emergence of resistance [36]. In particular, drugs targeting evolvability-modifying genes could be used in combination with antibiotics to block the most common mutational paths to resistance.

**Dependence of resistance evolution on the environment**
The success of any treatment strategy based on evolution experiments, however, depends on whether resistance evolved in the laboratory is comparable to that observed in the clinic. The growth media used in laboratory evolution typically allow for rapid growth to accelerate evolution, but are undoubtedly different from the mostly nutrient-poor environments bacteria encounter in infections, raising the question: To what extent are resistance mutations dependent on the environment? As many resistance mutations lead to slower growth in the absence of drugs [37], the most beneficial resistance mutations depend on antibiotic concentration and the nutrient environment [38,39]. Further, mutations in metabolic genes are often selected in resistance evolution, especially when the evolution protocol includes long drug-free periods [38]; such periods may occur in infection environments.

In a recent study, the link between antibiotic resistance and metabolism was understood from bacterial growth laws [39], of which the most prominent example is the linear relation between growth rate and intracellular ribosome concentration [41,42]. By constraining bacterial physiology, these empirical laws enable predictions of the nutrient-environment-dependent inhibition caused by ribosome-targeting antibiotics [43] and even drug–drug interactions [44]. Some resistance mutations not only reduce drug uptake but also inhibit nutrient uptake, exemplifying the link between metabolism and drug resistance. Coupling this link with the growth laws enabled predictions of the growth rate of different resistant mutants and the predominant resistance mechanism selected at different drug and nutrient levels [39].

While these examples highlight the potential of system-level mathematical models of antibiotic resistance evolution, this approach is currently limited to translation inhibitors [39,43,44]. It would be interesting to extend it to other drugs such as antibiotics targeting DNA replication [45]; to this end, one would need to identify relationships similar to the established growth laws for antibiotic targets other than the ribosome – assuming they exist. Our understanding of bacterial physiology has benefited greatly from systematically varying the environment [46], but such studies are lacking for resistance evolution. Finally, most research on resistance evolution has focused on single-strain communities, whereas it is becoming increasingly clear that ecological interactions among species play a key role in antibiotic resistance [47,48].

**Acknowledgments**
We thank Gerrit Ansmann and Theresa Fink for their feedback on the manuscript. This work was supported in part by German Research Foundation (DFG) Collaborative Research Centre (SFB) 1310 and by a research fellowship of the Alexander von Humboldt Foundation (to YM).

**Box 1. Antibiotic resistance evolution assays.**
Technical aspects of evolution experiments can strongly affect their outcome. Particularly relevant is the control of the drug concentration, which determines the selection pressure during the experiment.
**a) Selection at constant antibiotic concentration.**

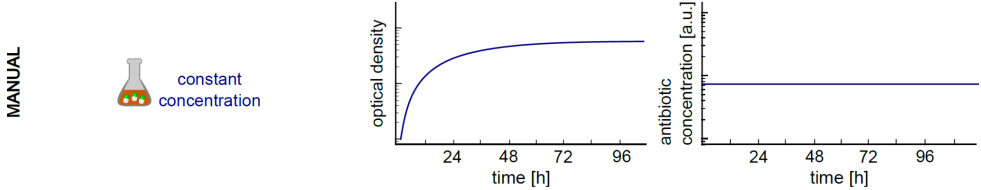

The simplest approach is to expose the bacterial population to a constant concentration throughout the experiment [25,27]. As a single point mutation can increase drug resistance several-fold, one round of selection usually largely eliminates the selection pressure; the population may also die out if the initial drug concentration is too high.

**b) Serial transfer with increasing antibiotic concentration.**

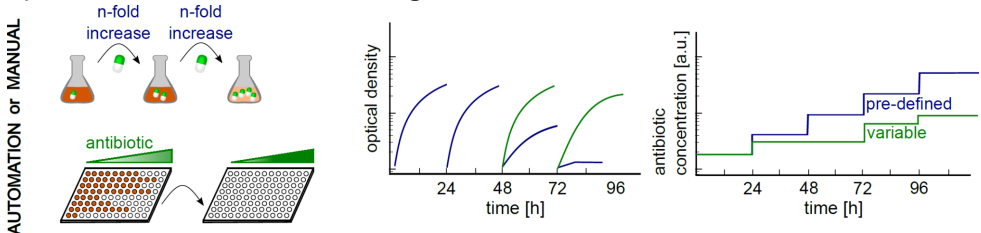

Multiple rounds of selection are achieved by increasing the drug concentration during the experiment, e.g. every 24 h [23,24,26]. Several studies have used a fixed fold-increase in drug concentration (blue lines) [23,49]. However, the frequency and size of the concentration steps strongly affects the observed results: If the predefined step size is too large, populations that evolve too slowly will die out; others will evolve resistance at the predefined rate, but no faster. This limitation can be overcome by dividing the population among several drug concentrations at each transfer and then continuing the experiment from the maximum concentration exhibiting growth (green lines) [26,38,50]. Laboratory automation platforms can drastically increase the throughput of these approaches [26,32].

**c) Feedback-controlled selection pressure and population size.**

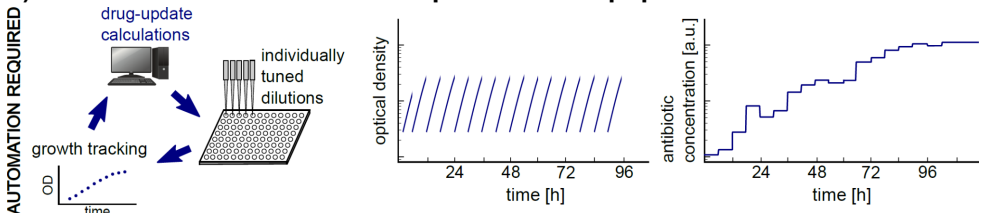

To maintain a tightly controlled selection pressure, adjustments in drug concentration need to be made at least every few hours, in flexible step sizes calculated for each population depending on a real-time estimate of its current resistance [5,24]. Frequent dilutions also maintain exponential growth and enable simultaneous control of population size, which is critical in evolution. This approach was first implemented in a custom-built continuous-culture device, the "morbidostat" [5]. An integrated robotic system was recently used to implement this feedback approach in high throughput [24].

Different evolution protocols underlie some notable differences in the results reviewed in this article. In particular, constant exponential growth with rapid updating of selection pressure typically leads to hundredfold increases in resistance within a week [5,24], compared with typical increases of less than tenfold achieved with other approaches [25,26,40]. This may also explain why no diminishing-returns trend was observed in several studies [25,26] – resistance may need to approach its saturation level for this trend to become apparent.

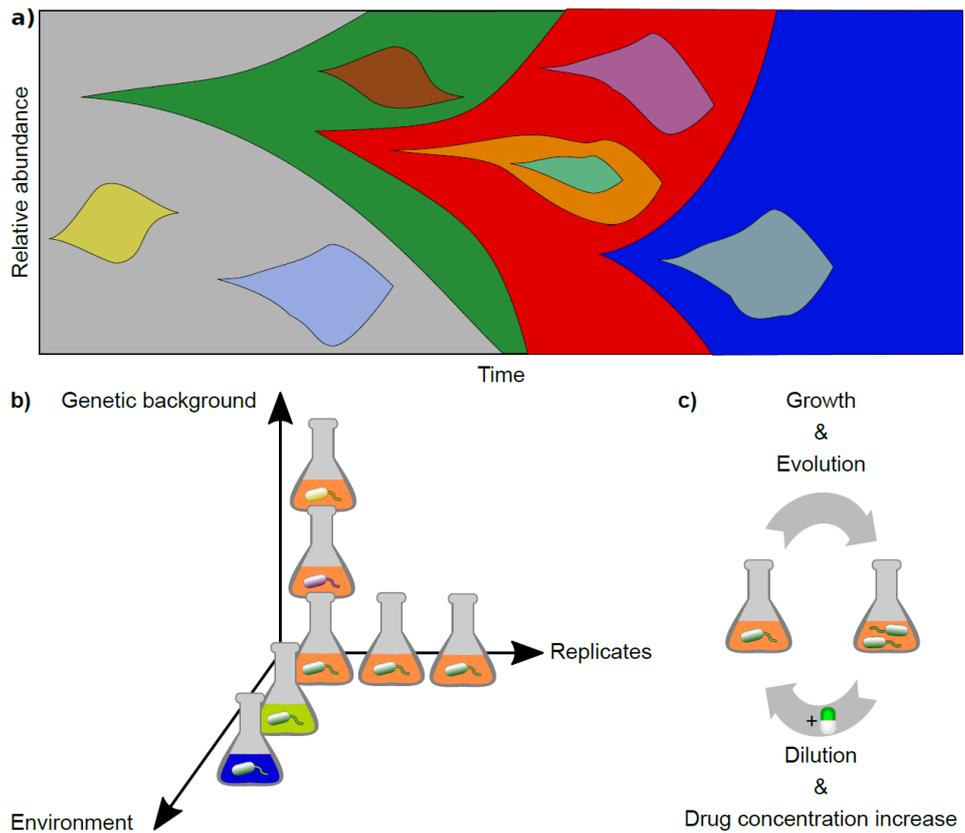

**Figure 1. Challenges in the study of antibiotic resistance evolution.
a)** Hypothetical Muller diagram showing how different lineages (different colors) arise over time from an initially clonal population (grey). **b)** Evolution experiments require high degrees of parallelism not only for replicate experiments (horizontal axis), but also to assess the effect of different environments (differently colored media, perpendicular axis) and genetic backgrounds (differently colored bacteria, vertical axis). **c)** As bacteria grow and evolve drug resistance (top arrow), dilution and an increase of the drug concentration are required to maintain growth and selection pressure (bottom arrow; *cf.* Box 1).